\documentclass[journal=jacsat,manuscript=article]{achemso}

\usepackage[version=3]{mhchem} 
\usepackage[T1]{fontenc}       
\usepackage{graphics}
\usepackage[usenames, dvipsnames]{color}
\usepackage{longtable}
\usepackage{multirow}
\usepackage{makecell}
\usepackage{natmove}
\setkeys{acs}{usetitle = true}
\usepackage{amsmath,achemso}
\usepackage{bm}
\setkeys{acs}{maxauthors = 9}

\author{Duy Khanh Nguyen}
\affiliation {Department of Physics, National Cheng Kung University,  Tainan, Taiwan}

\author{Ngoc Thanh Thuy Tran}
\affiliation {Hierarchical Green-Energy Materials (Hi-GEM) Research Center,
National Cheng Kung University, Tainan, Taiwan}

\author{Yu-Huang Chiu}
\affiliation {Department of Applied Physics, National Pingtung University,  Pingtung, Taiwan}

\author{Ming-Fa Lin}
\email{mflin@mail.ncku.edu.tw}
\affiliation {Physics department/QTC/Hi-GEM, National Cheng Kung University,  Tainan, Taiwan}

\title[An \textsf{achemso} demo]
  {Concentration-Diversified Magnetic and Electronic Properties of Halogen-Adsorbed Silicene}

\abbreviations{DOS, XPS, STM, ARPES, STS}

\begin{document}


\begin{abstract}

 Diverse magnetic and electronic properties of halogen-adsorbed silicene are investigated by the first-principles theoretical framework, including the adatom-diversified geometric structures, the atom-dominated energy bands, the spatial spin density distributions, the spatial charge density distributions and its variations, and the spin- and orbital-projected density of states. Also, such physical quantities are sufficient to identify similar and different features in the double-side and single-side adsorptions. The former belongs to the concentration-depended finite gap semiconductors or p-type metals, while the latter display the valence energy bands with/without spin-splitting intersecting with the Fermi level. Both adsorption types show the halogen-related weakly dispersed bands at deep energies, the adatom-modified middle-energy \(\sigma\) bands,  and the recovery of low-energy \(\pi\)  bands during the destruction of the halogen concentrations.  Such feature-rich band structures can be verified by the angle-resolved photoemission spectroscopy experiment.

\end{abstract}

\section{1. Introduction}

A new era of  material sciences has  arisen since the  successful fabrication of two-dimensional (2D) graphene\cite{1,2}. Such a 2D planar carbon structure with a hexagonal lattice exhibits many exceptional properties, including  massless Dirac fermions, strength of the lattice structure, high thermal conductivity, and half-integer Hall conductance\cite{3,4,5}. However, the lack of an energy gap makes graphene incompatible for practical applications. Therefore, how to induce a band gap  is an important issue to make use of graphene as a real material and  provides a strong motivation for studies on graphene composites/graphene-like 2D materials. The graphene composite materials with a sizable band gap consisting of grapheneoxide (GO)\cite{6,7}, graphane (HO)\cite{8,9,10}, fluorographene (CF)\cite{11,12}, and chlorographene (CCl)\cite{13,14} have investigated in previous studies. Also, the graphene-like 2D monolayers of BN are explored in a stable structure\cite{15,16}. Such a BN system is a wide gap insulator with an energy gap of 4.6 eV even though BN has the same 2D planar honeycomb structure as graphene\cite{17}. Therefore, the tiny lattice mismatch in the graphene/BN contact renders it possible to construct nanoscale electronic devices\cite{18}. Furthermore, graphene-like 2D materials comprised of group IV elements, which possess a buckled honeycomb structure with a mix of  sp\(^2\)/sp\(^3\) hybridization formed by the four outermost-orbital valence electrons, contain a variety of potential properties and have attracted tremendous attention in the fields of physics, chemistry, and material science, especially the low buckled 2D silicon structures\cite{19,20,21,22,23}.

 Though the premature allotrope forms of silicon, nanotubes\cite{24} and fullerene\cite{25}, were early  identified,  a 2D silicon structure does not seem to exist in nature owing to the absence of a solid phase of silicon. As a result, monolayer silicene is impossible to synthesize by exfoliation methods as initially utilized for graphene. Other possible methods are advanced for the growth of silicene. The most feasible method is to deposit silicon atoms on metal surfaces (silver or iridium) that do not interact strongly with Si atoms or create compounds\cite{26,27,28,29}, providing the direct evidence for the presence of 2D silicon sheets which were theoretically predicted in 1994\cite{30}. Formation energy and phonon dispersion calculations have determined that silicene is energetically favorable in a low buckled honeycomb structure\cite{31}.  The electronic structure of pristine silicene presents a negligible energy gap, in which the energy bands linearly cross at the Fermi level. This feature might attribute a massless Dirac fermion characteristic to charge carriers, which  has been verified by angle-resolved photoemission spectroscopy  and scanning tunneling spectroscopy  measurements\cite{32,33}. The possible high carrier mobility make silicene an ideal material, especially for the applications in field effect transistors\cite{34}. Further extraordinary properties of silicene have been revealed, such as a large energy gap  opened by the spin-orbit coupling at Dirac point\cite{35}, a quantum spin Hall effect that can be  observed in an experimentally accessible low-temperature regime\cite{36}, transition from a topological insulating phase to a band insulator which can be generated by an electric field and electrically tunable energy gap\cite{37,38}, and the emergence of a valley-polarized metal and anomalous quantum  Hall effect\cite{39}. Such potential properties identify silicene as a promising candidate to replace graphene not only due to the graphene-like features, but also its compatibility with  silicon-based electronic devices.

Most intriguingly, the essential properties of silicene are extremely sensitive to  chemical doping. Among the chemical functionalizations, the various adatom dopings are one of the effective methods to dramatically change the electronic structures. Hydrogen-adsorbed silicene exhibits semiconducting or metallic behavior that depend on the various adatom configurations with different concentrations\cite{40}. The metalization of silicene, narrow gap semiconductor, and the semi-metallic or semiconducting behavior are revealed in alkali (Li, Na, K)-, alkali-earth (Be, 	Mg, Ca)-, and 3d transition metal atom (Ti, V, Cr, Mn, Fe, Co, Mo, and W)-adsorbed silicenes, respectively\cite{41}. Besides,  the chemical functionalization of silicene with boron (B), nitrogen (N), aluminum (Al), and phosphorus (P) adatoms have been explored in detail.  Such systems present a metallic behavior with strongly bonded B, N, Al, and P adatoms, in which there exists an obvious electron transfer from Si atoms to B, N, and P adatoms (p-type doping), and the opposite charge transfer (n-type doping) with weaker bonding for Al adatoms\cite{42}. Furthermore, the electronic properties and optimal structure models of halogenated silicene have been explored by first-principles studies and low-temperature scanning tunneling microscopy  measurements, respectively\cite{43,44,45}. However, those studies mainly focused on  a specific halogen concentration. As a result, the various halogenation  effects on 2D silicene deserve a further look because of its promising  applications, especially for electrode materials in lithium-ion battery.

In this work, the geometric, magnetic, and electronic properties of halogen-adsorbed silicene are rigorously investigated by the first-principles study. The double-side [Fig. 1(a)] and single-side [Fig. 1(b)] adsorptions are considered in calculations to evidence that these two adsorption types can present the diverse magnetic and electronic  properties, in which they are determined by the adatom-diversified geometric structures, atom-dominated energy bands, the spatial spin density distributions, the spatial charge density distributions and its difference, and the spin- and orbital-projected density of states (DOSs).  Via such physical quantities, the similar and different characteristics in double-side and single-side adsorption cases are thoroughly identified, including the halogen-related weakly dispersed bands at deep energies, the adatom-modified middle-energy \(\sigma\) bands, the recovery of low-energy \(\pi\) bands during the destruction of the halogen concentrations; however, the breaking of the mirror symmetry in the single-side adsorptions leads to the critical discrepancy, i.e., there appear the valence energy bands with/without spin-splitting intersecting with the Fermi level as a result of p-type metallic behavior. These theoretical predictions can be verified by the scanning tunneling microscopy (STM), angle-resolved photoemission spectroscopy (ARPES), and scanning tunneling spectroscopy (STS) measurements.

\section{2. Computational method}

The diverse magnetic and electronic properties of halogen-adsorbed silicene are investigated by the spin-polarized density functional theory (DFT) implemented in VASP (Vienna ab initio simulation package)\cite{46}. The exchange and correlation energies, which come from the many-particle electron-electron interactions, are evaluated by the Perdew-Burke-Ernzerhof (PBE) functional under the generalized gradient approximation\cite{47}. Furthermore, the projector-augmented wave (PAW) pseudopotentials can characterize the electron-ion interactions. The plane-waves basis set with a kinetic energy  cutoff of 500 eV is utilized to calculate wave functions and state energies. A vacuum space of \( 10\)  \AA  \ is inserted between periodic planes to avoid the interaction of two planes. The first Brillouin zone is sampled by  \( 12 \times 12 \times 1 \)
   and \( 100 \times 100 \times 1 \) k-point meshes within the Monkhorst-Pack scheme for geometric optimizations and electronic structure calculations, respectively. Such points are sufficient for obtaining reliable DOSs.  The convergence for energy is set to be 
   \(10^{ - 5}   \)
   eV between two consecutive steps, and the maximum Hellmann-Feynman force acting on each atom is less than \(  0.01  \)
    eV/\AA \, during the ionic relaxations.

\section{3. Result and discussion}
\subsection{3.1. Geometric structure}

The DFT calculations are performed in order to explore the geometric, magnetic, and electronic properties of halogen-adsorbed silicene under the various concentrations.  Two typical adsorption types of fully halogenated silicene are chosen for this systematic study. The double-side adsorption, in which both sides of the silicene plane are thoroughly covered by halogen adatoms, as shown in Fig. 1(a). The halogen adatoms are uniformly distributed so that each adatom is separated by its six nearest adatoms with the same distance [Fig. 1(c)], and the back-side  adatoms (green balls) are set in the middle of three front-side adatoms (red balls). For the single-side adsorption, the adatoms only cover a single side of the silicene plane [Fig. 1(b)], i.e., all the back-side adatoms are removed so that the  concentration  is reduced to half of the double-side adsorption. The  typical concentrations vary from  100\( \%\), 12,5\( \%\), and  5,6\( \%\), where the ratios between  halogen adatoms and Si atoms correspond to 2:2, 1:8 and 1:18, respectively [Fig. 1(c)];  more cases are shown in Tables 1 and 2.  The binding energy is calculated by $E_b$ = ($E_{T}$ - $E_{P}$ - n$E_{A}$)/n, in which the $E_{T}$, $E_{P}$, and $E_{A}$  are, respectively, the energies of the fully halogen adsorbed silicene, pristine silicene, and isolated halogen adatoms; n represents the number of halogen adatoms. The lower binding energy corresponds to a higher stability. The most optimal geometric symmetry is used in the calculations. Via the detailed examinations, the most optimal adsorption position is situated at the top site (on the top of the upper silicon atom), as compared with the hollow site (above the center of the hexagonal silicon ring), valley site (on the top of the lower silicon atom), and bridge site (on the top of the Si-Si bond), irrespective of any  doping case\cite{41}. Among the halogen adatoms, the magnitude of binding energy declines with the increase of the atomic number [Tables 1 and 2], i.e., the F-adsorbed systems achieve the lowest binding energy ${\sim\,-5.3}$ eV - \(- 4.7\) eV. Furthermore, the halogen-absorbed silicene [Tables 1 and 2] possesses a higher geometric stability than halogen-adsorbed graphene\cite{48} because of its highly reactive buckled surface.  The shortest halogen-Si bond length is revealed in the F-adsorbed system ${\sim\,1.63}$ \AA \  - ${1.64}$ \AA \ , consistent to its smallest atomic number among halogen atoms. The stable halogen-Si  bondings are formed by the transfer of electrons from Si atoms to halogen adatoms (discussed in the charge distribution section), and they are almost insensitive to the adatom concentration. The charge transfer renders the nearest Si-Si bonds weaker, whereas the second-nearest Si-Si bonds become stronger, i.e., the bond lengths for the nearest and second-nearest Si-Si bonds are lengthened (${2.31}$ \AA \ - ${2.33}$ \AA \ ) and shortened (${2.23}$ \AA \ - ${2.25}$ \AA \ ), respectively.   Besides, the halogen-Si-Si bond angles increase and buckling structures become obvious when the adatom concentration  declines [Tables 1 and 2].  It is notable that in a pristine silicene, the planar  Si-Si bondings are formed by Si-(3s, 3p\(_x\), 3p\(_y\)) orbitals, while the parallel 3p\(_z\) orbitals of nearest Si atoms  are reduced the \(\pi\)-\(\pi\) overlap, and thus Si atoms have a mix of sp\(^2\)/sp\(^3\) hybridization; therefore, the adatom adsorption on silicene could be easier to tune than purely sp\(^2\) hybridized graphene.  When the halogen adatoms are bonded with Si atoms, the Si atoms are changed from a mix of sp\(^2\)/sp\(^3\) hybridization to  sp\(^3\) hybridization.  The chemical bonding scheme is determined by the hybridization between the Si-3p\(_z\) orbitals and halogen-(p\(_x\),  p\(_y\), and p\(_z\)) orbitals, and the weak sp\(^3\) hybridization of four orbitals (3s, 3p\(_x\), 3p\(_y\), and 3p\(_z\)). The double-side and single-side adsorptions present similar geometric distortions; however, they might lead to different electronic and magnetic properties, mainly owing to the broken mirror symmetry [Fig. 1(b)].

For experimental verification, STM is a powerful instrument for imaging surfaces at the atomic level. This tool can directly detect the surface structures in real-space under the atomic resolution, including the atomic lattice, the very short bond lengths, the planar or buckled structures, the achiral or chiral edges, the surface adsorptions, and the direct substitutions. Up to date, high-resolution STM observations have successfully verified the atomic structures of graphenes\cite{49}, few-layer graphenes\cite{50}, hydrogenation of graphenes\cite{51}, fluorinated graphenes on copper\cite{52},  1D graphene nanoribbons\cite{53},  silicene layers grown on Ag(111)\cite{27},  silicene nanoribbons on Ag(110)\cite{54}, hydrogenation of silicene films grown on Ag(111)\cite{55}, chlorination of monolayer silicenes\cite{45}, and multilayer silicenes \cite{56}. Apparently, the predicted results in the buckled silicene after halogen adsorptions, including the top-site positions of halogen adatoms,  halogen-Si $\&$ Si-Si bond lengths, buckling structures, and bond angles could be verified by  high-resolution STM  measurements, being very useful in confirming the multi-orbital hybridizations in critical chemical bonds.

\subsection{3.2.  Rich band structure}
     
The 2D energy dispersions along the highly symmetric points (\(\Gamma \)-K-M-\(\Gamma \)) provide much useful information in  examining the main characteristics of the electronic properties. These energy dispersions reveal a dramatic change under the competition between the halogen-Si bondings and the weak sp\(^3\) hybridization. A monolayer silicene exhibits a Dirac cone structure at the K point (corner of the hexagonal Brillouin zone) owing to the extended \(\pi\) bondings of Si-3p\(_z\) orbitals, and a negligible energy gap of \(1\) meV.  The low-lying linear energy bands turn into parabolic dispersions with a saddle point at the M point, as shown in Fig. 2(a). These energy bands are mainly dominated by 3p\(_z\) orbitals of the two nearest Si atoms, regarding as \(\pi\) bands.  Moreover, the weak sp\(^3\) hybridization illustrates an obvious separation of the \(\pi\) bands and \(\sigma\) bands within \(\pm 2\) eV. The occupied and unoccupied \(\sigma\) parabolic bands are respectively initiated at \(-1\) eV and \(1.2\) eV. The highest occupied bands display a double degeneration at the \(\Gamma \) point, while the lowest unoccupied bands are non-degenerate at the M point. The former bands are mainly contributed by the (3p\(_x\) and 3p\(_y\)) orbitals, and they are hybridized with the 3p\(_z\) orbitals at lower energy of \(-2.5\) eV.  The third occupied \(\sigma\) band initiates at ${-1}$ eV, dominating by the 3s orbitals, and gradually becomes parabolically dispersed.  This band becomes partially flat at \(-3.2\) eV  near the \(\Gamma \) point, in which there exist an orbital hybridization between the 3s orbitals and 3p\(_z\) orbitals (discussed in DOSs section). These fundamental features are dramatically modified after halogen adsorptions.

  The double-side adsorption cases exhibit semiconducting  or metallic behaviors, depending on distinct concentrations.  The fully fluorinated, chlorinated, brominated, and iodinated silicene present a direct energy gap of \(0.47\) eV, \(1.16\) eV, \(1.09\) eV, and \(0.51\) eV,  as shown in  Figs. 2(b), 2(e), 2(f), 2(g), respectively. These gaps are determined by the highest occupied states (HOS) and lowest unoccupied states (LUS) at the \(\Gamma \) point, and their magnitudes do not have a linear relationship with their atomic numbers.  Otherwise, fully astatinated silicene displays a semi-metallic behavior with an overlap of valence and  conduction bands at the Fermi level near the \(\Gamma\) point owing to the very weak At-Si bonds [Fig. 2(h)].  The hybridization of Si-3p\(_z\) orbitals and halogen-p\(_z\) orbitals puts the \(\pi\) bands away from the Fermi level.  Thus, the band gaps are determined by the \(\sigma\) bands. The middle-energy \(\sigma\) band of Si-(3p\(_x\) and  3p\(_y\)) orbitals at \(-1\) eV [Fig. 2(a)] is situated ${\sim\,-0.2}$ eV [Fig. 2(b)] or ${\sim\,-0.5}$ eV [Figs. 2(e) and 2(f)],  meaning that such bands have a blue shift of ${\sim\,0.8}$ eV or ${\sim\,0.5}$ eV.  This clearly indicates the Si-Si \(\sigma\) bondings display a small change after halogen adsorptions, observing from the lengthened nearest Si-Si bond lengths (discussed in the geometric structure section) and the weakened nearest Si-Si bond strength (discussed in the spatial charge distribution section). Among halogen adatoms, F adatom-related weakly dispersed bands dominate at the deepest energies, due to its shortest F-Si bond length [blue circles in  Fig. 2(b)].  This band dispersion becomes more obvious and gradually dominates at higher energies with an increase of its atomic numbers, mainly owing to the higher halogen-Si bond lengths [blue circles Figs. 2(e), 2(f), 2(g), and 2(h)].

    As the adatom concentration declines, the halogen-Si bondings no longer dominate, but rather compete with the effect of the weak sp\(^3\) hybridization to determine the energy bands. Quite different from the \(100\%\) system, the \(25\%\) system becomes an indirect energy gap of \(0.75\) eV,  which is determined by HOS and LUS at the \(\Gamma \) and M points [Fig. 2(c) and Table 1]. The critical concentration is revealed at \(11\%\) [Fig. 2(d)],  exhibiting the p-type metallic behavior, in which there exist valence energy bands intersecting with the Fermi level, so that the unoccupied valence states between the Fermi level and the top of valence bands belong to free holes. Also, the energy bands near the Fermi level with a weaker dispersion are mainly contributed by the non-passivated Si-3p\(_z\) orbitals [Fig. 2(c) and 2(d)], revealing that the low-lying energy bands gradually changed from the \(\sigma\) bands to the \(\pi\) bands. Besides, the middle-energy  \(\sigma\) band of Si-(3p\(_x\) and  3p\(_y\)) orbitals in pristine silicene [Fig. 2(a)] is located  ${\sim\,-0.3}$ eV [Fig. 2(c)] and ${\sim\,-0.7}$ eV [Fig. 2(d)], reflecting a small variation in Si-Si  bondings under these low-concentration adsorptions. Furthermore, such middle-energy \(\sigma\) bands are contributed by both the passivated Si atoms and non-passivated Si atoms, in which the latter atoms contribute much to such-bands in the \(11\%\) systems because of its higher concentration.  Notably, the direct energy gap in the \(100\%\) system becomes indirect when the concentration reduces up to \(25\%\) [Table 1], and any halogen concentration lower than the critical concentration of \(11\%\) becomes the p-type metals.  This remains true for most of the halogen adatoms.  These revealed band features are sufficient to comprehend the concentration-dependent electronic properties.

In the single-side adsorptions, the destroy of the mirror symmetry creates the  valence energy bands with/without spin-splitting intersecting with the Fermi level, in which the Fermi level is situated at the valence band, so the unoccupied valence states between the Fermi level and the half-occupied valence energy bands belong to free holes [Fig. 3]. As a result, these systems can be regarded as the p-type metals.  The full single-side adsorption cases [Figs. 3(a), 3(d), 3(f), 3(g), and 3(h)] have valence parabolic bands with/without intersecting with Fermi level,  which are only related to the three non-passivated Si atoms [gray circles in Fig. 1(b)] nearest to the passivated Si atoms [red circles in Fig. 1(b)]. The parabolic dispersion change to the partially flat bands when the concentration declines, as shown in Figs. 3(b)/(e) and 3(c) for \(12.5\%\) and \(5.6\%\) systems, respectively.  The spin-splitting energy bands become more obvious near the Fermi level, in which the splitting spacing is about 1 eV in the \(50\%\) systems, as shown by red (spin up) and black (spin down) solid curves in Figs. 3(d), 3(f), 3(g), and 3(h). However, this spin-splitting spacing becomes about \(0.3\) eV  smaller in \(12.5\%\) systems [Figs. 3(b) and (c)] and vanishes at the critical concentration of the \(5.6\%\) [Fig. 3(c)]. The spin-splitting bands can be further confirmed by spatial spin density distributions [Fig. 4] and asymmetric peaks near the Fermi level in DOSs [Fig. 7]. Furthermore, the \(\pi\) bondings between non-passivated Si atoms and passivated Si atoms [Fig. 1(b)] is seriously suppressed owing to the high charge transfer from the latter Si atoms and the halogen adatoms. This affirms that the non-passivated Si atoms can not form the extended \(\pi\) electronic states via the interaction with the passivated Si atoms. The clear discrepancy between the single-side and double-side adsorption cases is the valence energy bands with/without intersecting with the Fermi level. Also, the halogen-induced weakly dispersed bands in the double-side adsorption cases [Fig. 2] are reduced their numbers in the single-side adsorption cases [Fig. 3] due to the halogen concentration decreased.

As compared with the pristine case [Fig. 2(a)],  the Dirac cone structure are thoroughly destroyed  in single-side and double-side adsorption cases, mainly owing to the termination or seriously destruction of the low-energy \(\pi\) bondings. Also, there exist other similar features, such as the recovery of low-energy \(\pi\) bands in the low-concentration systems, the adatom-modified middle-energy \(\sigma\) bands, and the halogen-induced weakly dispersed bands at deep energies. When the concentration decreases, both \(5.6\%\) single-side [Fig. 3(c)] and \(11\%\) double-side [Fig. 2(d)] systems show the low-lying \(\pi\) bands with large energy widths. These low-energy \(\pi\)  bands are mainly contributed by the non-passivated Si-3p\(_z\) orbitals, indicating that the \(\pi\) bondings  are recovered among the Si atoms.  The middle-energy \(\sigma\) bands (\(-0.2\) eV - \(-3.5\) eV) [Figs. 3(a)-3(c) and 2(b)-2(d)] are very similar to each other in terms of their band width, initiated energy, and orbital contribution, mainly owing to the hardly affected \(\sigma\) bonds. In addition, the halogen-related  weakly dispersed bands are confined at similar energies in both single-side and double-side systems since the stable halogen-Si bondings are insensitive with the halogen concentrations (discussed in geometric structure section). That is to conclude that the broken geometric symmetries hardly have an impact on the aforementioned similar features.

To date, ARPES is the unique experimental tool to verify the main features of  band structures. This experimental equipment can directly measure the direction, speed, and scattering process of valence electrons in the sample being studied. The high-resolution ARPES experiments have successfully verified the feature-rich band structures of  graphenes on Ir(111)\cite{57}, few-layer graphenes\cite{58}, halogenated graphenes\cite{59}, 1D graphene nanoribbons\cite{60}, monolayer silicenes\cite{61},  hydrogenated silicenes\cite{62}, and multilayer silicene nanoribbons\cite{63}.  Apparently,  similar ARPES measurements can fully generate to examine the feature-rich band structures after halogen adsorptions, including the destroy or recovery of low-energy \(\pi\) bands, the halogen-induced weakly dispersed bands at deep energies, and the valence energy bands intersecting with the Fermi level.  These detailed evaluations are very useful to determine the multi-orbital hybridizations in critical chemical bonds.

\subsection{3.3.  Spatial spin distribution}

   The spatial spin density distributions accompanied by the net magnetic moments can provide  further information in comprehending the spin-splitting energy bands. The fully dominated halogen-Si bondings in the double-side adsorption cases [Fig. 1(a)] thoroughly destroy the spatial spin orientations, so that their energy bands are highly degenerate or without spin-splitting. When the full domination of halogen-Si bondings no longer exists [Fig. 1(b)], the spatial spin density distributions can be performed in the single-side adsorption cases, leading to the spin-splitting energy bands.  Only except for the \(50\%\) fluorinated system [Fig. 3(a)] exhibits the degenerate parabolic band.  The other \(50\%\) systems have the spin-splitting parabolic bands, which can clearly observe near the Fermi level, as shown by red (spin up) and black (spin down) curves in Figs. 3(d), 3(f), 3(g), and 3(h). These spin-splitting parabolic bands are mainly contributed by the non-passivated Si atoms, verifying from the spatial spin density distributions near the non-passivated Si atoms (pure spin up configuration) [Figs. 4(a), 4(b), and 4(c)] and highly asymmetric prominent peaks near the Fermi level in DOSs [Figs. 7(d), 7(f), 7(g), and 7(h)]. However, the spin-splitting bands become weaker in the \(12.5\%\) systems [Figs. 3(b) and 3(e)]. This indicates that the spin-up and spin-down states can co-exist in its spin density distributions as the halogen concentration declined (spin up and spin down mixed configuration), as shown by red (spin up state) and blue (spin down state) balls in Figs. 4(d), 4(e), and 4(f). The difference in the spin up and spin down states directly determines the strength of the net magnetic moments,  i.e., \(0.72\) \( \mu _B \), \(0.89\) \( \mu _B \), and \(0.79\) \( \mu _B \) in the \(50\%\) systems  and \(0.64\) \( \mu _B \), \(0.67\) \( \mu _B \), and  \(0.45\) \( \mu _B \) in the \(12.5\%\) systems [Table 2].  Notably, the spin density distributions are fully absence at the critical concentration of \(5.6\%\) (vanishing the net magnetic moment in Table 2), thus its energy bands are highly degenerate [Fig. 3(c)]. Furthermore, any halogen concentration beyond the critical concentration results in the absence of the spin density distributions.  This remains true for all halogen adatoms.  The single-side adsorption-induced magnetic properties on the silicene surface could be verified using spin-polarized STM.
   
 \subsection{3.4.  Spatial charge distribution}

The spatial charge density distribution ($\rho$)  provides  very useful information on the chemical bondings of all orbitals and thus elucidate the dramatic change of energy bands.  Obviously, $\rho$  can illustrate the  strength of the Si-Si and halogen-Si bondings, as shown in Figs. 5(a)-5(g).  \(\rho\) displays a strong  \(\sigma\) bonding between two Si atoms in the pristine silicene [Fig. 5(a)] or two non-passivated Si atoms in the low-concentration systems [Figs. 5(f) and 5(g)], as presented by the enclosed black rectangle. The strength of such \(\sigma\) bonding becomes weaker when Si atoms are bonded to halogen adatoms, as indicated by enclosed gray rectangle in Figs. 5(b)-5(g), resulting in the extended Si-Si bond length. This accounts for the similarity in the adatom-modified  middle-energy \(\sigma\) bands between the single-side and double-side adsorption cases. Away from the horizontal line between two Si atoms, $\rho$ is lowered; however, its charge distribution is extended, evidencing for the \(\pi\)-electronic states, as shown by enclosed red rectangle in Fig. 5(a), 5(f), and 5(g). Nevertheless, hardly extended states, corresponding to the \(\pi\)-charge-depleted region, are perceived between the passivated and non-passivated Si atoms (enclosed purple rectangle in Figs. 5(d)-5(g)). This charge distribution is similar to isolated Si atoms and evidences for the destroy of \(\pi\) bondings.  Furthermore, the charge density of halogen-Si bondings is much higher than that of Si-Si bondings, as indicated by the dashed black rectangle in Figs. 5(b)-(g). Thus, such stable halogen-Si bondings are insensitive to the halogen concentrations. This illustrates that the halogen-Si bonding-induced weakly dispersed bands are limited to similar energies in both single-side and double-side systems.  In order to further understand the detailed charge transfers among all orbitals, the charge density difference (${\Delta\rho}$) is presented in Figs. 5(h)-5(m).  ${\Delta\rho}$ is created by subtracting the charge density of the pristine silicene and isolated halogen adatoms from that of the composite system.  All the halogen-adsorbed systems exhibit a high charge transfer from Si atoms to halogen adatoms to form the stable halogen-Si bondings (dashed black rectangle in Figs. 5(h)-(m)). Also, such clear variation of charge density accounts for a weaker bond strength of the nearest Si-Si bonds around the halogen adatoms.

\subsection{3.4.  Diverse density of states}

The spin- and orbital-projected DOSs is very useful to affirm the main features in the band structures and comprehend the orbital hybridizations in critical chemical bondings. For pristine silicene [Fig. 6(a)], the low-energy DOSs exhibits the vanished value at \(E_F=0\),  V-shaped structure, and two symmetric logarithmic divergent peaks at ${E^{c,v}\sim\,\pm 1}$ eV, are induced by Si-3p\(_z\) orbitals (purple solid curve in Fig. 6(a)).  They respectively originate from the negligible-energy gap, the Dirac cone structure, and the saddle points of \(\pi\) bands at the M point [Fig. 2(a)].  The middle-energy DOSs shows a shoulder structure at $-1$ eV and a prominent peak at ${-2.6}$ eV, resulting from the maximum point and the saddle point of \(\sigma\) bands, respectively. They are mainly contributed by Si-(3p\(_x\) and 3p\(_y\)) orbitals (green and blue solid curves in Fig. 6(a)). Moreover, the hybridization between Si-3s (red solid curve in Fig. 6(a)) and Si-3p\(_z\) orbitals creates a prominent peak at ${-3.2}$ eV, coming from the partially flat band with an initiated energy at the \(\Gamma\)  point.  These features further  indicate that the interactions between \(\pi\) and  \(\sigma\) bands at low energy are weak.

The low-energy DOSs is strongly modified after halogen adsorptions. Under the full double-side adsorption cases,  it shows the vacant states at specific region centered at \(E_F=0\) in the fluorinated [Fig. 6(b)], chlorinated [Fig. 6(e)], brominated [Fig. 6(f)], and iodinated systems [Fig. 6(g)], coming from the finite gap feature in its corresponding band structures. Otherwise, the finite value of DOSs at \(E_F=0\) comes to exist in the astatinated system [Fig. 6(h)] as result of the semi-metallic behavior [Fig. 2(h)].  The middle-energy DOSs displays a shoulder structure at ${-0.2}$ eV and the prominent peak at ${-1.5}$ eV, are dominated by Si-(3p\(_x\) and  3p\(_y\)) orbitals (green and blue solid curves in Fig. 6(b)), are very similar to that of pristine silicene [Fig. 6(a)], accounting for the adatom-modified middle-energy \(\sigma\) band. Such modified middle-energy DOSs structure is also observed in Figs. 6(e)-6(h).   Furthermore, the extra strong peaks at deep energies are dominated by halogen-(2p\(_x\) and 2p\(_y\)) orbitals [Fig. 6(b)], standing for the halogen-related weakly dispersed bands at deep energies [Fig. 2(b)]. These strong peaks become weaker and dominate at higher energies as their atomic numbers increasing [Figs. 6(e)-6(h)]. When the halogen concentration declines, three low-energy prominent peaks  appear at \(-0.6\) eV, \(0.8\) eV, and \(1.4\) eV  in the \(25\%\) system [Fig. 6(c)]. The bandwidth of these prominent peaks about ${0.6}$ eV, and their states are related to Si-3p\(_z\) orbitals (purple solid curve). Also, their middle-energy DOSs are contributed by Si-(3p\(_x\) and 3p\(_y\)) orbitals; however, the single peak structure in the full adsorption case [Fig. 6(b)] separate into several sub-peaks owing to an enhancement of sub-bands. At the critical concentration of \(11\%\), most of the low-energy DOSs within \(\pm 1\) eV are mainly dominated by the non-passivated Si-3p\(_z\) orbitals [Fig. 6(d)], illustrating a \(\pi\) bandwidth similar to the pristine case [Fig. 6(a)]. This evidences the recovery of low-energy \(\pi\) bands in low-concentration systems.  Also, these two low-concentration systems show the halogen-dominated extra strong peaks at deep energies similar to the \(100\%\)  system [Fig. 6(b)].  However, the intensity of these deep-energy extra strong peaks is roughly proportional to the halogen concentration.

The fundamental features of DOSs in the single-side and double-side adsorption cases are the strong peaks, which are closely associated with the Si- and halogen-related bondings. The similarities are identified between the single-side cases and its corresponding double-side adsorptions, including the extended low-energy  \(\pi\) band widths [Figs. 7(c) and 6(d)], the three Si-3p\(_z\) orbital-induced prominent peaks within \(-1\) eV  to \(2\) eV [Fig. 7(b) and Fig. 6(c)], the Si-(3p\(_x\) and 3p\(_y\))  orbitals-dominated middle-energy single-peak structure [Fig. 7(a) and Fig. 6(b)], and the halogen-dominated extra strong peaks at deep energies [Figs. 7(a) and 6(b); Figs. 7(b) and 6(c); Figs. 7(c) and 6(d)].  These extra strong peaks are confined at the same energy range; however,  their intensities are gradually declined as the halogen concentration reduced. The similar DOSs features is also found in other halogen-adsorption cases [Figs. 7(d-h) and 6(e-h)].  On the other hand, the breaking of the mirror symmetry in the single-side adsorption cases leads to a critical discrepancy, i.e., there exist the finite value of DOSs at the Fermi level as a result of the metallic behaviors [Figs. 7(a)-7(h)]. Furthermore, the asymmetric prominent peaks near the Fermi level [Figs. 7(b), 7(d-h)] are the useful evidence in confirming the spin-splitting energy bands [Fig. 3].

  On the experimental side, the STS measurements, where the tunneling differential conductance (dI/dV) is proportional to DOSs, can provide  sufficient information in examining the special features in DOSs. High-resolution STS measurements are able to distinguish the semiconducting, semi-metallic and metallic behavior. Moreover, they can be used to identify the close relations between the electronic energy spectra and the orbital hybridizations of the chemical bonds.  To date, such experimental measurements have been successfully used to confirm the electronic band structure near the Fermi level and the dimension-diversified van Hove singularities in  monolayer graphene\cite{64} and few-layer graphene systems\cite{65,66,67}, 1D graphene nanoribbons \cite{68,69}, adatom-adsorbed graphene\cite{70}, monolayer silicene\cite{71}, and hydrogenated silicene\cite{72}.  Obviously, the theoretical predictions on the halogen-diversified DOSs can be thoroughly affirmed by the STS experiments, including the vanishing or existing of the finite value of DOSs at the Fermi level, the Si- or halogen-dominated strong peaks, and the halogen-modified \(\sigma\) shoulder structures. Furthermore,  spin-polarized STS is available for verifying the asymmetric peaks near the Fermi level in DOSs.

\section{4. Conclusion}

The geometric, magnetic, and electronic properties of halogen-adsorbed silicene are studied by DFT calculations. The rich physical and chemical properties are diversified by the halogen concentrations. Specifically, the stable halogen-Si bondings are the critical factor in affecting the geometric structures, spatial spin density distributions, spatial charge density distributions and its variations, atom-dominated energy bands, and spin- and orbital-projected DOSs. The significant transfer of electrons from Si atoms to halogen adatoms leads to variations in the charge density distributions in critical chemical bondings (different bond lengths), directly resulting in the drastically modulation of the magnetic and electronic properties. The rich band structures after halogen adsorptions are highlighted by the destruction or recovery of low-lying \(\pi\) bands, the halogen-induced weakly dispersed bands at deep energies, and the valence energy bands with/without spin-splitting intersecting with the Fermi level.  The first feature is determined by the competition between the halogen-Si bondings and the weak sp\(^3\) hybridization. The energy gap determined by the \(\sigma\) bands is indirect in the full double-side systems. As the concentration declines, the low-lying \(\pi\) bands gradually recover, rendering the gap size smaller with indirect behavior in the \(25\%\) system, and even though becoming the p-type metals at the critical concentration of \(11\%\).  The second feature, mainly owing to the stable halogen-Si bonding insensitive with the halogen concentrations, is the similar characteristic of both single-side and double-side adsorptions. The third feature, originating from the broken mirror symmetry, is a distinct characteristic in the single-side adsorptions. Furthermore, spatial spin arrangements only reveal in the single-side systems since the full domination of halogen-Si bondings longer exist.  The above-mentioned band properties are confirmed by their DOSs, including the vacant region centered at the Fermi level, the recovery of \(\pi\) band-related low-energy prominent peaks,  the halogen-related extra strong peaks at deep energies, and the finite value of DOSs at the Fermi level. The feature-rich band structures and the diverse DOSs can be verified by the ARPES and STS measurements, respectively. The concentration-tuned diverse magnetic and electronic properties of halogen-adsorbed silicene are very potential for applications in both nanoelectronics and spintronics.
 
 \par\noindent {\bf Conflict of interest}
   
   There are no conflicts of interest in this paper

  \par\noindent {\bf Acknowledgments}
     
 This work was supported by the Physics Division, National Center for Theoretical Sciences (South), the Nation Science Council of Taiwan under the grant No. NSC-102-2112-M-006-007-MY3.  This work was also supported in part by the Ministry of Science and Technology of Taiwan, Republic of China, under Grant No. MOST 107-2112-M-153-002.


\begin{table}[htb]
                    \caption{Binding energy (\(E_b\)) eV; magnetic moment/magnetism; energy gap (\( E_g^{d(i)} \))/metal (M)/semi-metal (SM); halogen (adatom)-Si bond length (\AA), Si-Si bond length (\AA),	adatom-Si-Si bond angle (\( ^\circ  \)), and buckling (\( \Delta \)) (\AA) for the double-side adsorption under the various concentrations. NM and FM correspond to non-magnetism and ferro-magnetism, respectively.}
                         \label{table2}
                           \begin{center}
                                                
                           \begin{tabular}{llllllllll}
                                \hline
                                & \makecell{Ratio of \\ adatom \\and Si}  & \(E_b\) (eV)
                                                 & \makecell{Magnetic\\ moment\\(\(\mu _B \) )/\\magnetism } & \makecell{ \( E_g^{d(i)} \) \\(eV)/ \\M/SM}  & \makecell{adatom\\-Si\\ (\AA)} & \makecell{Near-\\est\\Si-Si\\(\AA)}&  \makecell{
                                                 \(
                                                 2^{nd} 
                                                 \)
                                                 \\ near-\\est\\Si-Si\\(\AA)}& \makecell{adatom\\-Si\\-Si \\angle \\(\( ^\circ  \))} & \makecell{\( \Delta \)  \\ (\AA)}  \\ 
                                                \hline
                                                 & Pristine & X & 0/NM & \( E_g^{d}=0.001\) & X & 2.255 & 2.255 & X & 0.490 \\ 
                                                 & F:Si=2:2 & \(-5.29490\) & 0/NM & \( E_g^{d}=0.47\) & 1.631 & 2.322 & X &  109.12 & 0.760 \\ 
                                                 & F:Si=8:8 & \(-5.29490\) & 0/NM & \( E_g^{d}=0.47\) & 1.631 & 2.322 & X & 109.12 & 0.760 \\ 
                                             & F:Si=6:8 & \(-5.38383\) & 0/NM & \( E_g^{i}=0.81\) & 1.630 & 2.335 & X & 109.63 & 0.762 \\

                                                 & F:Si=4:8 & \(-5.32565\) & 0/NM & \( E_g^{i}=0.78\) & 1.630 & 2.332 & X & 109.67 & 0.763 \\

                                                 & F:Si=2:8 & \(-5.31450\) & 0/NM & \( E_g^{i}=0.75\) & 1.639 & 2.319 & 2.235 & 110.42 & 0.771 \\

                                                 & F:Si=2:18 & \(-5.39129\) & 0/NM & M & 1.637 & 2.318 & 2.234 & 111.92 & 0.887 \\ 
                                                 
                                             & F:Si=2:32 & \(-5.45832\) & 0/NM & M & 1.641 & 2.318 & 2.249 & 113.96 & 0.910\\             
                                          
                                                 & Cl:Si=2:2 & \(-3.22631\) & 0/NM & \( E_g^{d}=1.16\) & 2.071 & 2.328 & X & 109.52 & 0.778 \\ 
                                               
                                                & Cl:Si=8:8 & \(-3.22685\) & 0/NM & \( E_g^{d}=1.16\) & 2.071 & 2.328 & X & 109.52 & 0.778 \\ 
                                             & Cl:Si=4:8 & \(-3.29581\) & 0/NM & \( E_g^{i}=0.93\) & 2.087 & 2.330 & X & 109.69 & 0.782 \\                       
                                                        
                                   & Cl:Si=2:8 & \(-3.26759\) & 0/NM & \( E_g^{i}=0.83\) & 2.096 & 2.321 & 2.236 & 109.83 & 0.785 \\

                              & Cl:Si=2:18 & \(-3.38563\) & 0/NM & M & 2.096 & 2.322 & 2.234 & 111.74 & 0.881 \\                                    
                            & Cl:Si=2:32 & \(-3.39675\) & 0/NM &  M & 2.097 & 2.321 & 2.233 & 111.91 & 0.898 \\

                                                 & Br:Si=2:2 & \(-2.41261\) & 0/NM & \( E_g^{d}=1.09\) & 2.240 & 2.326 & X & 109.42 & 0.772 \\ 
                                                 
                                                 & Br:Si=8:8 & \(-2.41256\) & 0/NM & \( E_g^{d}=1.09\) & 2.240 & 2.326 & X & 109.42& 0.772 \\ 
                                  & Br:Si=4:8 & \(-2.75106\) & 0/NM & \( E_g^{i}=0.96\) & 2.261 &  2.325 & X & 109.76 & 0.780 \\                     
                                             
                                        & Br:Si=2:8 & \(-2.67977\) & 0/NM & \( E_g^{i}=0.94\) & 2.274 &  2.320 & 2.237 & 109.98 & 0.791 \\

                                   & Br:Si=2:18 & \(-2.79401\) & 0/NM & M & 2.274 & 2.320 & 2.233 & 111.54 & 0.864 \\                    
                                 & Br:Si=2:32 & \(-2.83164\) & 0/NM & M & 2.276 & 2.320 & 2.233 & 111.72 & 0.902 \\

                                                  & I:Si=2:2 & \(-1.29362\) & 0/NM & \( E_g^{d}=0.51\) & 2.459 & 2.326 & X & 109.41 & 0.772 \\ 
                                                  
                                                 & I:Si=8:8  & \(-1.29426\) & 0/NM & \( E_g^{d}=0.51\) & 2.459 & 2.326 & X &  109.41 & 0.772\\  
                                            & I:Si=4:8 & \(-2.00385\) & 0/NM & \( E_g^{i}=0.95\) & 2.485 & 2.313 & X & 110.63 & 0.785\\                      
                                          & I:Si=2:8 & \(-1.97356\) & 0/NM & \( E_g^{i}=0.93\) & 2.507 & 2.320 & 2.238 & 110.95 & 0.789 \\

                                                    & I:Si=2:18 & \(-2.19425\) & 0/NM & M & 2.506 & 2.319 & 2.233 & 111.67 & 0.854 \\    
                                            & I:Si=2:32 & \(-2.20716\) & 0/NM &  M & 2.508 & 2.320 & 2.233 & 111.92 & 0.893 \\

                                                 & At:Si=2:2  & \(-0.71032\)  & 0/NM & SM & 2.584 & 2.326 & X & 109.43 & 0.772 \\  
                                                  
                                             \hline

                                                \end{tabular}
                                                 \end{center}
                                                 \end{table}  

\begin{table}[htb]
                    \caption{Representation similar to that in Table 1, but shown for single-side adsorptions.}
                         \label{table2}
                           \begin{center}
                                                
                           \begin{tabular}{llllllllll}
                                \hline  & \makecell{Ratio of \\ adatom\\ and Si}  & \(E_b\) (eV)
                                                 & \makecell{Magnetic\\ moment\\(\(\mu _B \) )/\\magnetism } & \makecell{ \( E_g^{d(i)} \) \\(eV)/ \\M/SM}  & \makecell{adatom\\-Si\\ (\AA)} & \makecell{Near-\\est\\Si-Si\\(\AA)}&  \makecell{
                                                 \(
                                                 2^{nd} 
                                                 \)
                                                 \\ near-\\est\\Si-Si\\(\AA)}& \makecell{adatom\\-Si\\-Si \\angle \\(\( ^\circ  \))} & \makecell{\( \Delta \)  \\ (\AA)}  \\ 
                                       \hline
                                                & F:Si=1:2 & \(-4.76899\) & 0/NM & M & 1.633 & 2.321 & X & 109.04 & 0.763 \\ 
                                              
                                                                                                    & F:Si=4:8 & \(-4.76846\) & 0/NM & M & 1.633 & 2.321 & X & 109.04 & 0.763\\

                                                 & F:Si=1:8 & \(-5.12832\) & 0.45/FM & M & 1.636 & 2.315 & 2.254 & 114.24 & 0.949 \\

                                                 & F:Si=1:18 & \(-5.23563\) & 0/NM & M & 1.636 & 2.313 & 2.259 & 116.96 & 1.048 \\

                                                & F:Si=1:32 & \(-5.31026\) & 0/NM & M & 1.637 & 2.313 & 2.259 & 117.40 & 1.064 \\

                                                 & Cl:Si=1:2 & \(-2.58719\) & 0.72/FM & M & 2.091
                                                  & 2.328 & X & 109.56 & 0.779 \\ 
                                                   
                                     & Cl:Si=4:8 & \(-2.58354\) & 0.72/FM & M & 2.091
                                                              & 2.328 & X & 109.56 & 0.779 \\

                                                 & Cl:Si=1:8 & \(-3.18866\) & 0.64/FM & M & 2.096 & 2.320 & 2.254 & 113.96 & 0.942 \\

                                           & Cl:Si=1:18 & \(-3.28117\) & 0/NM & M & 2.099 & 2.320 & 2.259 & 117.12 & 1.057 \\                         
                                        & Cl:Si=1:32 & \(-3.35039\) & 0/NM & M & 2.099 & 2.321 & 2.260 & 117.70 & 1.079 \\

                                              
                                                 & Br:Si=1:2 & \(-1.78324\) & 0.89/FM & M & 2.269 & 2.326 & X & 109.43 & 0.773 \\ 
                                                                               
                                              & Br:Si=4:8 & \(-1.78228\) & 0.89/FM & M & 2.269 & 2.326 & X & 109.43 & 0.773 \\

                                                 & Br:Si=1:8 & \(-2.55506\) & 0.67/FM	& M & 2.277 & 2.320 & 2.254 & 113.69 & 0.932 \\

                                                 & Br:Si=1:18 & \(-2.62592\) & 0/NM & M & 2.280 & 2.320 & 2.258 & 116.68 & 1.042 \\

                                                    & Br:Si=1:32 & \(-2.67667\) & 0/NM & M & 2.280 & 2.321 & 2.260 & 117.63 & 1.077 \\

                                                  & I:Si=1:2 & \(-0.83202\) & 0.79/FM & M & 2.566 & 2.336 & X & 110.08 & 0.802 \\
                                                  
                                                   & I:Si=4:8 & \(-0.83172\) & 0.79/FM & M & 2.566 & 2.336 & X & 110.08 & 0.802 \\

                                                  & I:Si=1:8 & \(-1.96195\) & 0.73/FM & M & 2.512 & 2.319 & 2.254 & 113.22 & 0.913 \\

                                                     & I:Si=1:18 & \(-2.07592\) & 0/NM & M & 2.517 & 2.320 & 2.258 & 116.45 & 1.033 \\          
                                            & I:Si=1:32 & \(-2.08183\) & 0/NM & M & 2.517 & 2.322 & 2.259 & 117.42 & 1.070 \\                            
                                                  
                                                  & At:Si=1:2 & \(-0.40616\) & 0.82/FM & M & 2.830 & 2.319 & X & 108.87 & 0.751 \\   
                                          
                                                \hline                                               
                                           \end{tabular}
                                                 \end{center}
                                                 \end{table}

\newpage
\begin{figure}[!h]
\centering
\includegraphics[width=12cm, height=20cm]{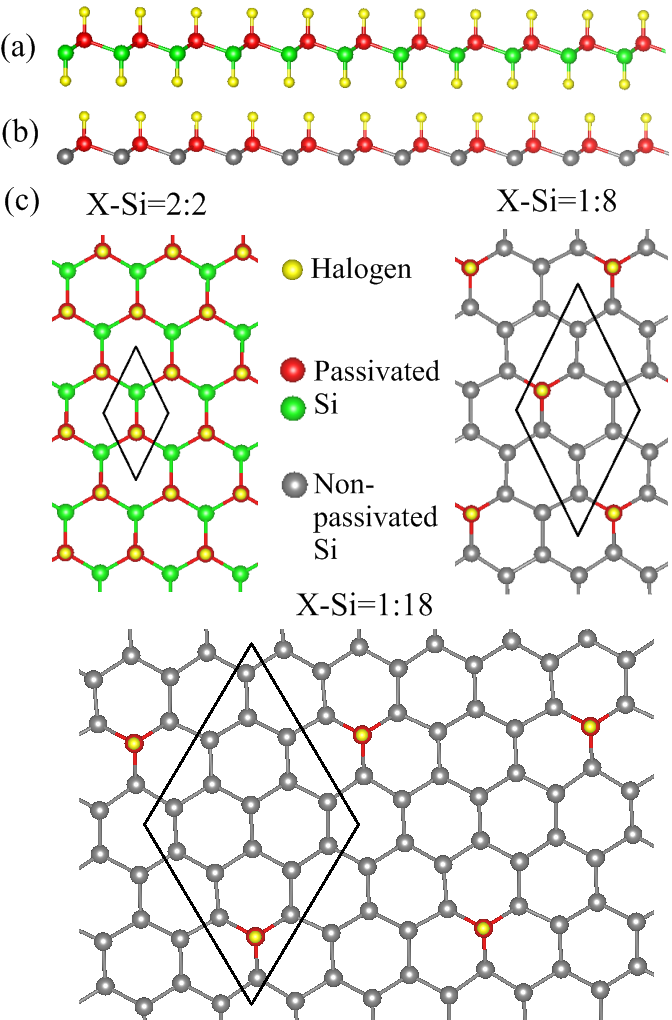}
\caption{Side-view geometric structures of halogenated silicene for (a) double-side adsorption and (b) single-side adsorption. (c) Top-view geometric structures of various halogen (X) concentrations, including X-Si=2:2, X-Si=1:8, and X-Si=1:18.}
\end{figure}

\newpage
\begin{figure}[!h]
\centering
\includegraphics[width=10cm, height=20cm]{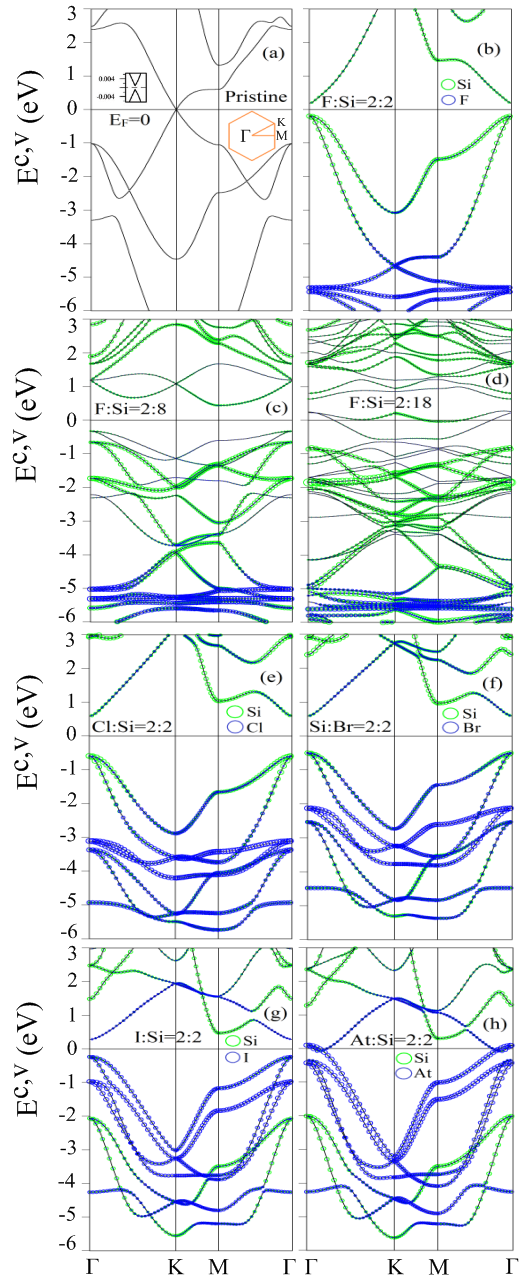}
\caption{Electronic band structures of (a) pristine silicene and halogenated silicene under double-side adsorptions for (b) F:Si=2:2, (c) F:Si=2:8, (d) F:Si=2:18, (e) Cl:Si=2:2, (f) Br:Si=2:2,  (g) I:Si=2:2, and (h) At:Si=2:2. Green and blue circles represent the contribution of passivated Si atoms and halogen adatoms, respectively.}
\end{figure}

\newpage
\begin{figure}[!h]
\centering
\includegraphics[width=10cm, height=20cm]{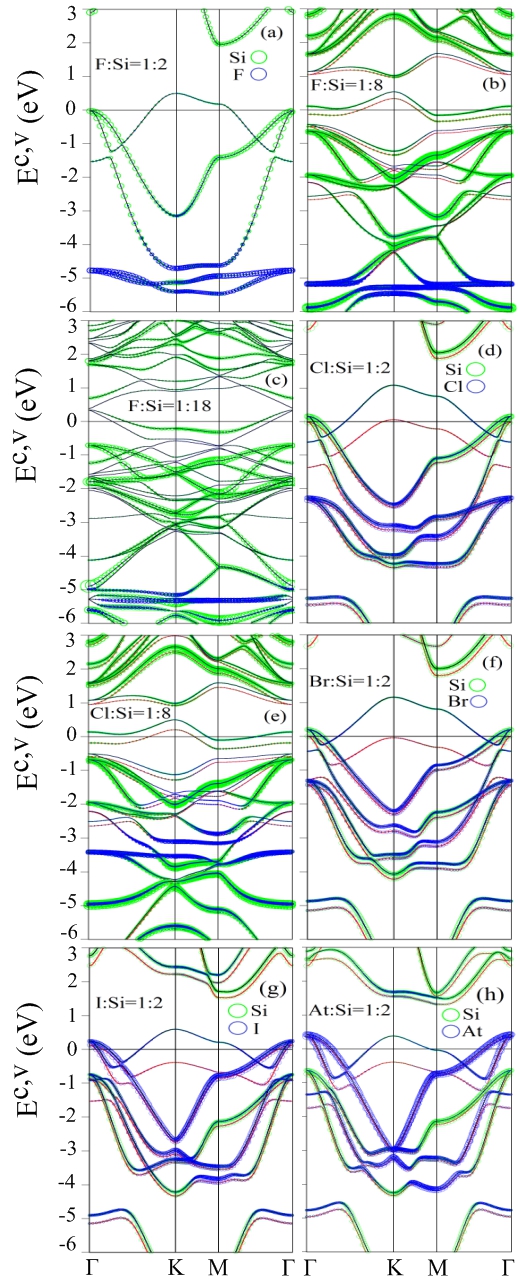}
\caption{Similar plot as Fig.2, but shown for  single-side adsorptions of (a) F:Si=1:2, (b) F:Si=1:8, (c) F:Si=1:18, (d) Cl:Si=1:2, (e) Cl:Si=1:8,  (f) Br:Si=1:2, (g) I:Si=1:2, and (h) At:Si=1:2. Red and black curves, respectively, correspond to the spin-up and spin-down energy bands.}
\end{figure}

\newpage
\begin{figure}[!h]
\centering
\includegraphics[width=12cm, height=17cm]{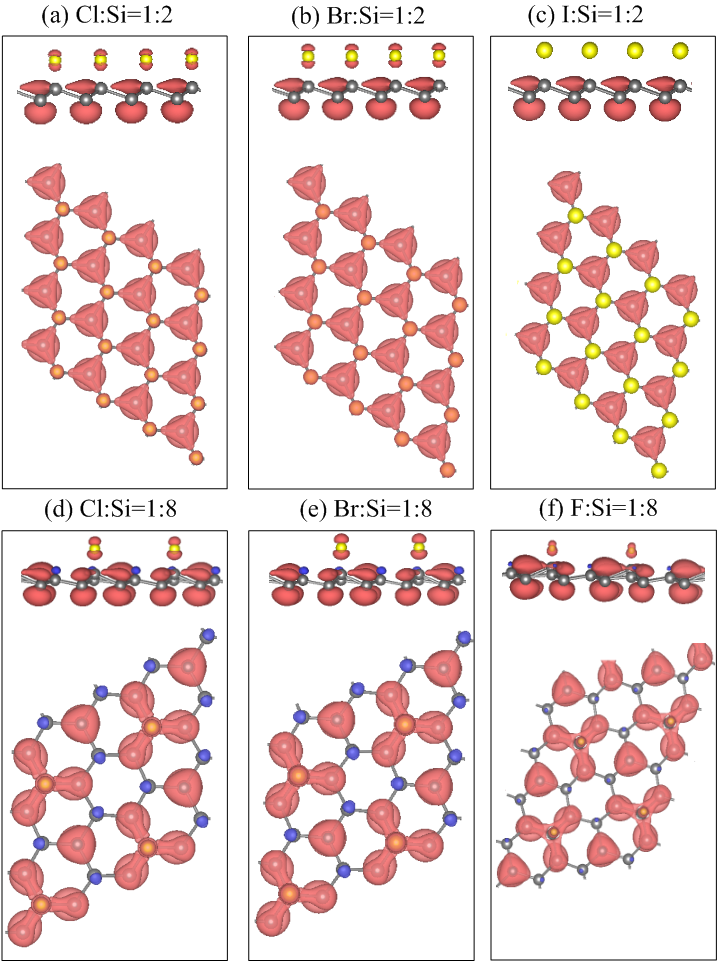}
\caption{Spatial spin density distribution with side-view and top-view for (a) Cl:Si=1:2, (b) Br:Si=1:2, (c) I:Si=1:2, (d) Cl:Si=1:8, (e) Br:Si=1:8, and (f) F:Si=1:8.  Red and blue isosurfaces, respectively, correspond to the spin-up and spin-down orientations.}
\end{figure}

\newpage
\begin{figure}[htb]
                \includegraphics[width=10cm, height=20cm]{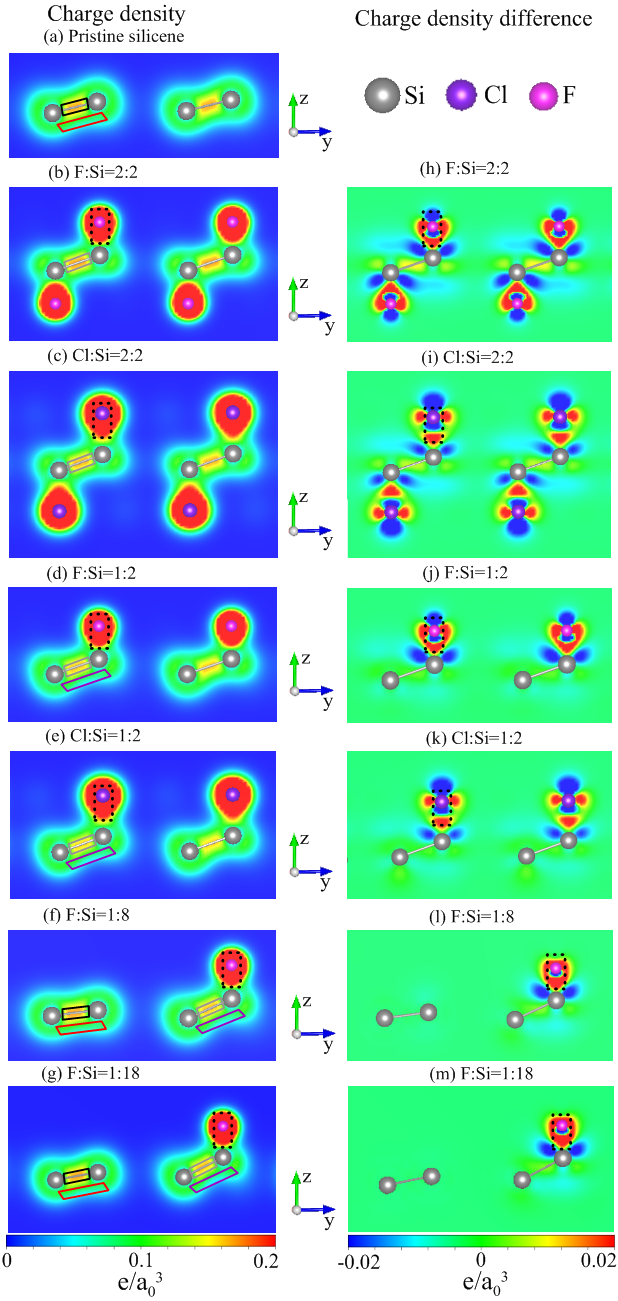}
                \caption{Spatial charge density at the left-hand side for (a) pristine silicene and halogenated silicene of (b) F:Si=2:2, (c) Cl:Si=2:2, (d) F:Si=1:2, (e) Cl:Si=1:2, (f) F:Si=1:8,  and (g) F:Si=1:18; charge density difference at the right-hand side  for (h) F:Si=2:2, (i) Cl:Si=2:2, (j) F:Si=1:2, (k) Cl:Si=1:2, (l) F:Si=1:8,  and (m) F:Si=1:18. $a_0$ is Bohr radius.}
                \label{fgr:6}
              \end{figure}

\newpage
\begin{figure}[htb]
                          \includegraphics[width=10cm, height=17cm]{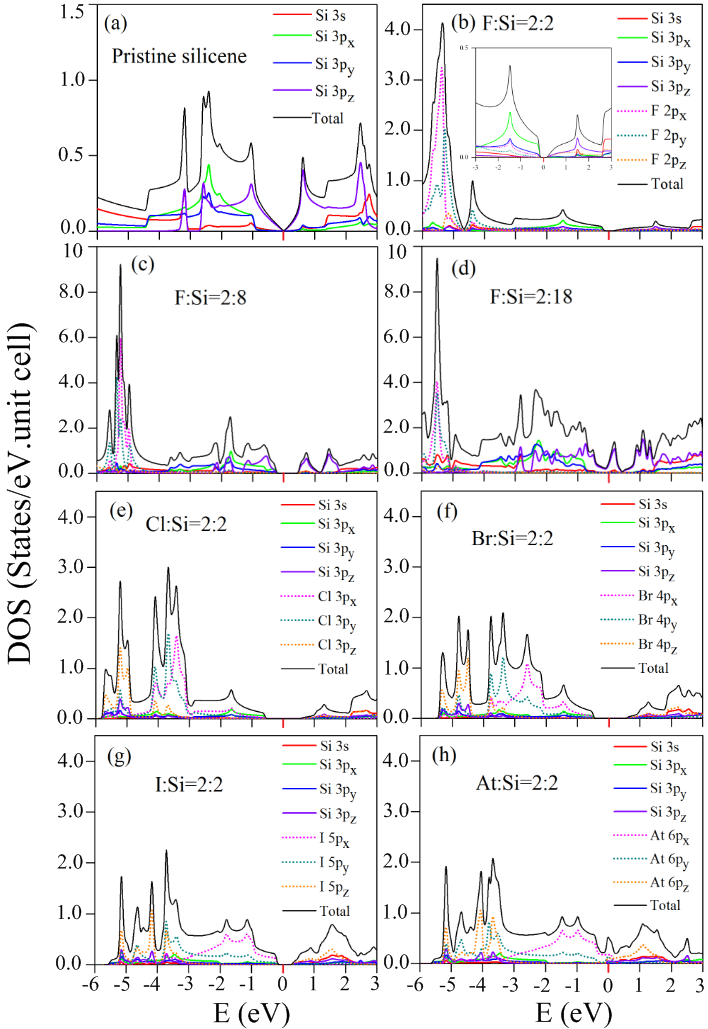}
                                     \caption{DOSs for (a) pristine silicene;  double-side adsorptions of (b) F:Si=2:2, (c) F:Si=2:8, (d) F:Si=2:18, (e) Cl:Si=2:2, (f) Br:Si=2:2;  (g) I:Si=2:2, and (h) At:Si=2:2.}
                                     \label{fgr:7}
                               \end{figure} 
                               
     \newpage
     \begin{figure}[htb]
                               \includegraphics[width=10cm, height=17cm]{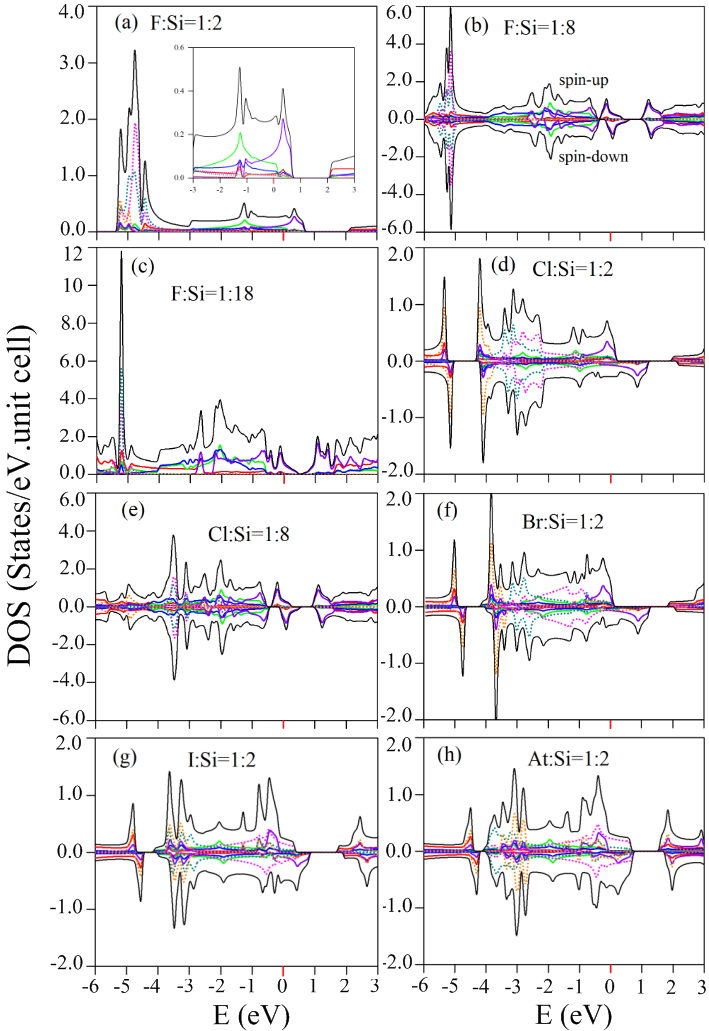}
                                          \caption{Similar plot as Fig. 6, but shown for  single-side adsorptions of (a) F:Si=1:2, (b) F:Si=1:8, (c) F:Si=1:18, (d) Cl:Si=1:2, (e) Cl:Si=1:8;  (f) Br:Si=1:2,  (g) I:Si=1:2, and (h) At:Si=1:2.}
                                          \label{fgr:7}
                                    \end{figure}                           
\end{document}